# Automatic Calibration Method of Voxel Size for Cone-beam 3D-CT Scanning System

**Min Yang(杨民)[1], Xiaolong Wang(王晓龙)[1], Yipeng Liu(刘义鹏)[2], Fanyong Men(孟凡勇)[3], Xingdong Li(李兴东)[4], Wenli Liu(刘文丽)[4], Dongbo Wei(魏东波)[1]**

1、School of Mechanical Engineering and Automation, Beijing Univ. of Aeronautics and Astronautics, Beijng 100191，China
2、Department of Control Science and Engineering, Harbin Institute of Technology, Harbin 150001, China
3、State Key Laboratory of Multiphase Complex Systems，Institute of Process Engineering, Chinese Academy of Science，Beijing 100190, China
4、National Institute of Metrology, Beijing 100013, China

**Abstract:** For cone-beam three-dimensional computed tomography (3D-CT) scanning system, voxel size is an important indicator to guarantee the accuracy of data analysis and feature measurement based on 3D-CT images. Meanwhile, the voxel size changes with the movement of the rotary table along X-ray direction. In order to realize the automatic calibration of the voxel size, a new easily-implemented method is proposed. According to this method, several projections of a spherical phantom are captured at different imaging positions and the corresponding voxel size values are calculated by non-linear least square fitting. Through these interpolation values, a linear equation is obtained, which reflects the relationship between the rotary table displacement distance from its nominal zero position and the voxel size. Finally, the linear equation is imported into the calibration module of the 3D-CT scanning system, and when the rotary table is moving along X-ray direction, the accurate value of the voxel size is dynamically exported. The experimental results prove that this method meets the requirements of the actual CT scanning system, and has virtues of easy implementation and high accuracy.

# 1. Introduction

With the rapid development of computer technology and the wide use of flat panel detector, three-dimensional CT(3D-CT) has recently generated intense interest from both scientific studies and practical applications in the non-destructive testing (NDT) field [1,2]. 3D reconstruction is carried out to get the internal structure and density distribution of the tested object through DR projection sequences collected by the detector at different views during the rotation of the tested object [3-6]. As one important step of 3D-CT, back projection addresses are calculated in the image geometry coordinate system with the unit of pixel, so the reconstructed data can not reflect the actual size of the tested object directly. For 3D-CT reconstruction, the tested object is considered as a virtual cuboid consisting of a large number of small cubes with a certain size stacked together regularly. These small cubes are named as voxels and their actual physical size is called voxel size, which is an important indicator to describe the resolution of the cone-beam 3D-CT scanning system. The smaller voxel size indicates the higher spatial resolution of a 3D image. The accurate value of the voxel size also guarantees the accuracy of data analysis and feature measurement based on 3D-CT images [7,8]. However, 3D reconstruction is realized in image geometry coordinate system with the unit of pixel, so the original voxel size must be in pixel unit. Accordingly, after the reconstruction, translating the original voxel size with pixel unit to the actual physical voxel size is a necessary step among the calibration of the cone beam 3D-CT scanning system. Furthermore, during CT scanning, when the tested object is placed in different positions between X-ray source and flat panel detector, the voxel size will be changed correspondingly. Therefore, we need to dynamically calibrate the voxel size complying with the variation of geometrical magnification ratio (GMR) caused by the alteration of the tested object's imaging position. The existing calibration methods need a special phantom to be scanned first, then the corresponding GMR and voxel size are calibrated. These methods are complicated and time-consuming, because when GMR changes, the phantom has to be scanned again [9,10]. Here we propose a new easily-implemented method to calibrate voxel size. According to this method, several projections of a spherical phantom are captured in different imaging positions and the corresponding voxel size values are calculated by non-linear least square fitting. Through these interpolation values, a linear equation is obtained, which reflects the rotary table displacement distance from its nominal zero position and the voxel size, so as to realize dynamical calibration of the voxel size. The experimental results prove that this method meets the requirements of the actual CT scanning system, and has virtues of easy implementation and high accuracy.

## 2. Methods

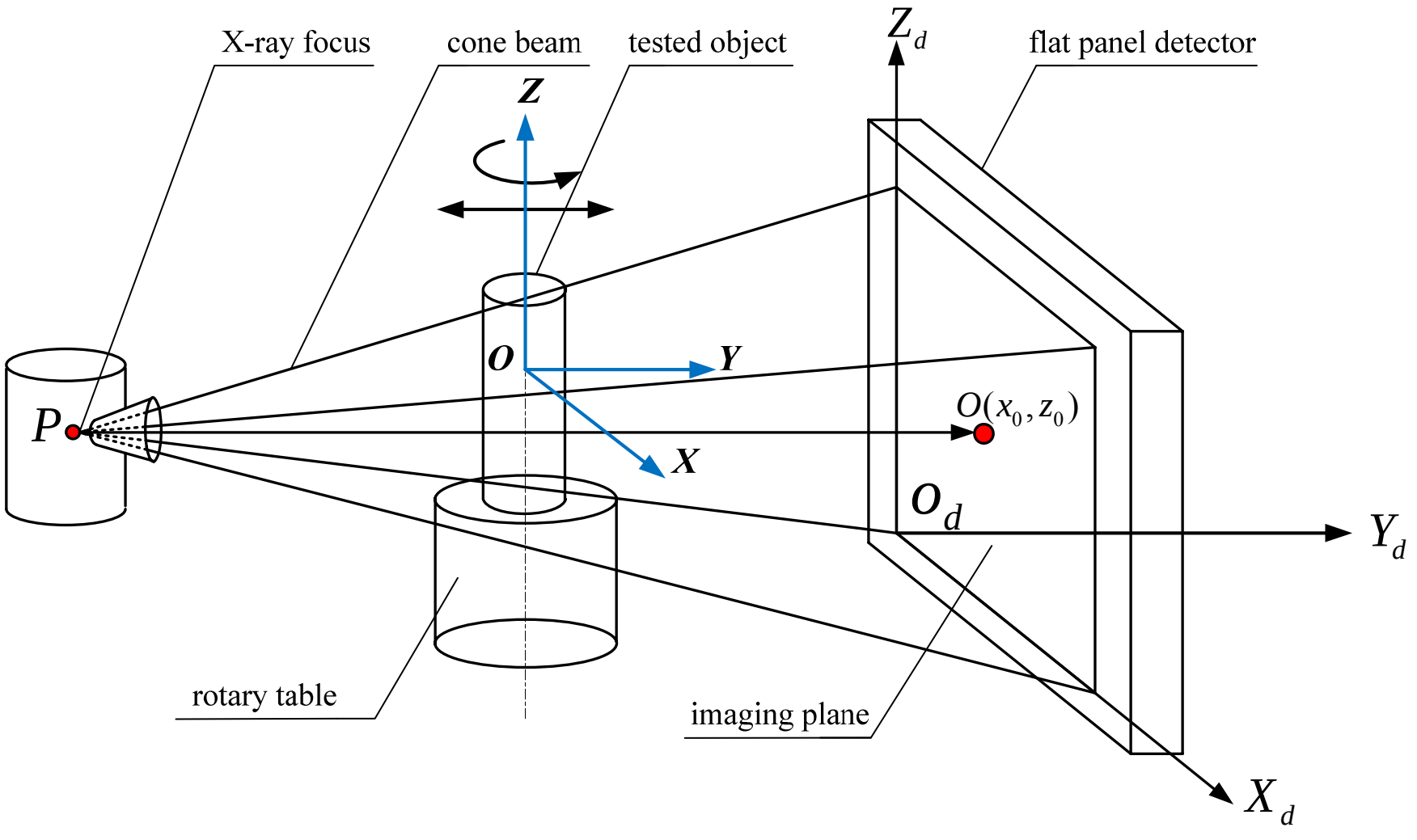

**Fig.1 The sketch of 3D-CT scanning ways**

The principle of a cone-beam 3D-CT scanning system is shown in Fig. 1. The cone-beam X-ray penetrates the tested object fixed on the rotary table and reaches the flat panel detector. The tested object rotates a certain angle step with the drive of the rotary table. The flat panel detector collects DR projections of the tested object at different imaging views among 360 degrees. With all DR projections, 3D reconstruction is performed and cross-sectional images are obtained. As shown in Fig.1, the imaging coordinate system is $X_d Y_d Z_d$, where 3D reconstruction operation is performed with the unit of pixel, and when the position of the detector in the imaging system is fixed, the imaging coordinate system will be determined. The object coordinate system is $XYZ$ where the actual size of the tested object lies. So before the feature measurement and data analysis through 3D-CT images, mapping the coordinate system $X_d Y_d Z_d$ to $XYZ$ is necessary, namely calculating the voxel size. Furthermore, when the tested object is moving between the X-ray source and the flat panel detector, the value of the voxel size will be changed correspondingly, so the automatic calibration is also needed. For 3D-CT scanning system, when the type of flat panel detector is determined, the physical size of the detection unit, namely the physical size of one pixel is constant, here we define the constant value as $U$. When one voxel unit in the tested object is projected to one pixel on the imaging plane of the detector, its projection size will be amplified by GMR times. Accordingly, the relationship between voxel size and detector unit size is:

$$V_i = U / M_i \quad (1)$$

where $M_i$ represents GMR at any imaging position, $V_i$ is the voxel size at $M_i$ imaging position. From the geometrical configuration in Fig.2, GMR at any imaging position can be calculated as:

$$M_i = D_{FOD} / L_i \qquad (2)$$

where $D_{FOD}$ is the distance from X-ray focus to detector imaging plane. When a 3D-CT scanning system is set up, the value of $D_{FOD}$ will be constant and can be measured by multi-projection methods [11,12]. $L_i$ is the distance from the rotation center of the rotary table to the X-ray focus, which cannot be measured directly. Obviously, $0 < L_i < D_{FOD}$ . Therefore, the difficulty of the calibration of $M_i$ is how to obtain the value of $L_i$ in formula (2) quickly and accurately.

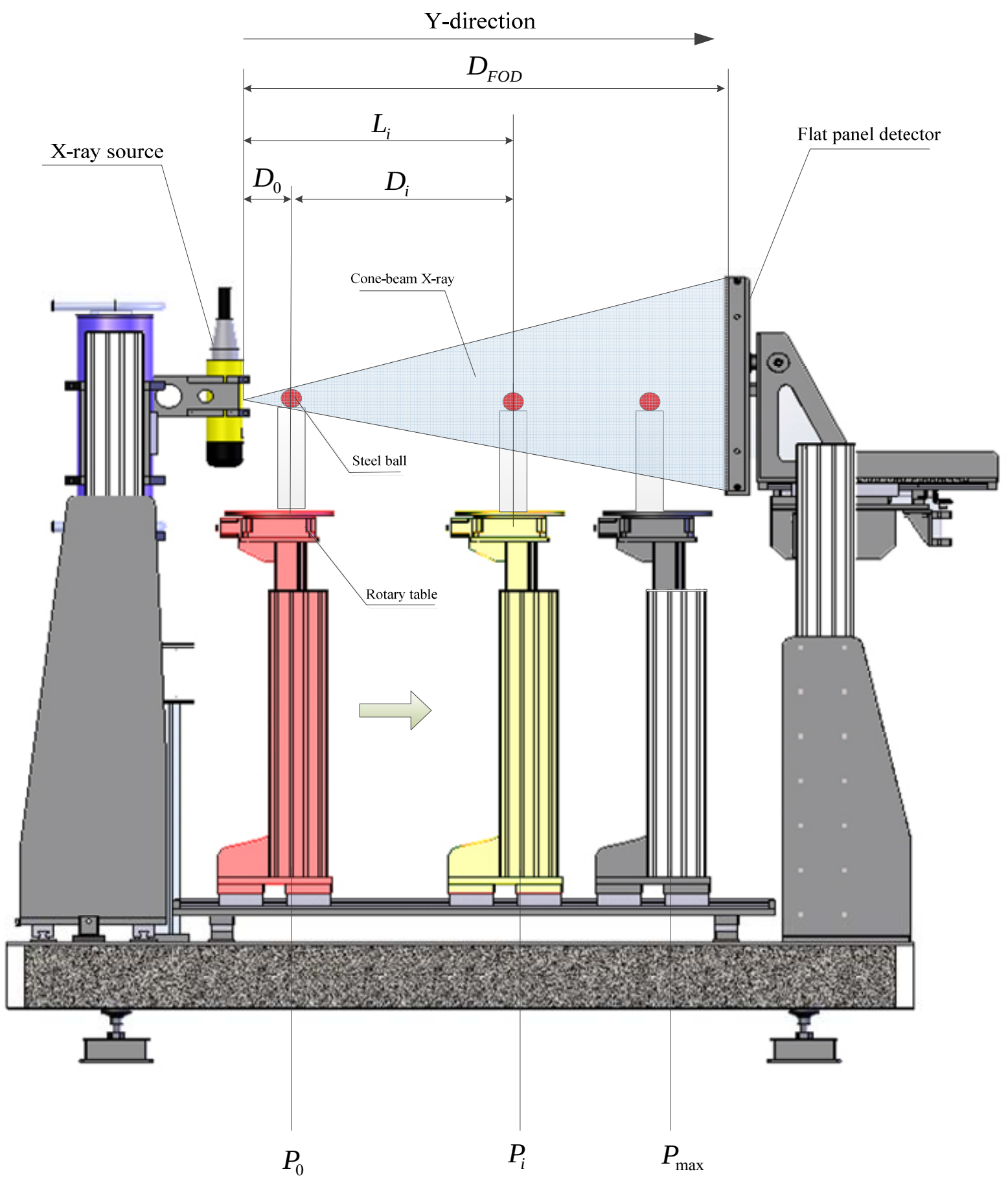

**Fig.2 The principle diagram of voxel size measurement**

From Fig. 2 , $L_i$ equals to

$$L_i = D_0 + D_i \tag{3}$$

where $D_0$ is the distance from the nominal zero position (also starting position) of the rotary table to the X-ray focus along Y-direction. The nominal zero position of the rotary table at Y-direction is $P_0$, which is designed well in advance according to the system movement control requirements. Thus after the 3D-CT scanning system is assembled, $D_0$ is also a constant value, which cannot be accurately measured unfortunately. When the rotary table moves to $P_i$ position within the range of movement along Y-direction, its translating distance from the starting position $P_0$ can be accurately recorded by motor controller, here we define the distance as $D_i$. Combining equation (1), (2) and (3), we have:

$$V_i = U\frac{D_0}{D_{FOD}} + \frac{U}{D_{FOD}} D_i \tag{4}$$

Let $b = U\frac{D_0}{D_{FOD}}$, $k = \frac{U}{D_{FOD}}$, obviously, $b$ and $k$ are constant values, then equation (4) becomes:

$$V_i = b + kD_i \tag{5}$$

The general expression of the relationship between voxel size and movement distance of the rotary table is:

$$V = b + kD \tag{6}$$

where $V$ represents the voxel size, and $V_i$ represents a sampled value of $V$. $D$ is the movement distance of the rotary table along Y-direction and $D_i$ represents a sampled value of $D$.

From equation (6), we can find that if we get the accurate values of $b$ and $k$, the automatic calibration of the voxel size is realized. Obviously, equation (6) is a linear equation, so the optimal way to get accurate values of $b$ and $k$ is linear fitting. According to the linear fitting method, some discrete values of $V$ and $D$ should be known in advance. In the practical implementation, we use a spherical phantom to realize linear fitting. As shown in Fig.2, first we get several DR (digital radiography) projections of the spherical phantom at different imaging positions along Y-direction, where the corresponding moving distances are recorded by motor controller, named as $[D_1, D_2, ......D_n]$. The projection is in circular shape. Then the image segmentation, edge detection, image thinning and moment match tracing are applied to each circular projection to obtain one closed circle contour of the projection [13,14]. During this step, a sub-pixel accuracy edge tracing method is applied to obtain the contour points, as shown in Fig.3, which is realized by two kinds of edge detection algorithms. First, the pixel-level detection by Canny operator is adopted, which has single-pixel response and good performance for improving the signal-to-noise ratio and location accuracy. Then, the moment match tracing algorithm is employed based on the pixel-level results. The coordinates of contour points with sub-pixel accuracy are acquired by solving the equations derived by an assumption that the ideal mixed moment is equal to the actual mixed moment of the image [15]. Based on the coordinates of contour points, the least square fitting algorithm is used to calculate the diameter of the circular projection.

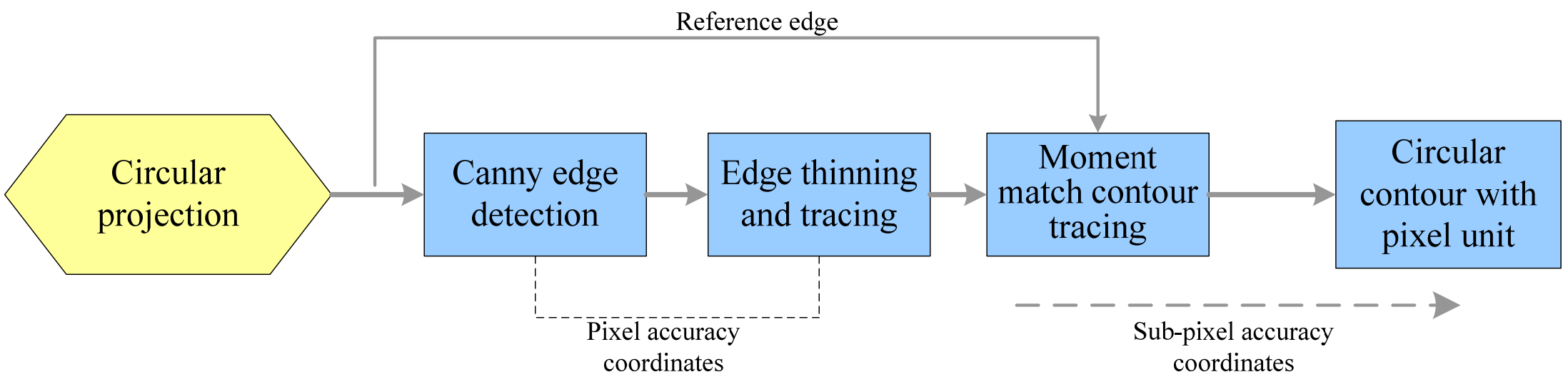

**Fig.3 Flow chart of sub-pixel accuracy edge tracing method**

The equation of the fitted circle is given as:

$$(x-z_0)^2+(y-z_0)^2=r^2 \qquad (7)$$

where $(x_0, z_0)$ is the center of the fitted circle, $r$ is the radius, and $(x_i, z_i)$ is the contour point of the circular projection. The error function of the fitted circle can be expressed as:

$$E=\sum_{i=1}^{n}(x_i^2-2x_ix_0+z_i^2-2z_iz_0+k)^2 \qquad (8)$$

where $k = x_0^2 + z_0^2 - r^2$

Let $\frac{\partial E}{\partial x_0}=0$、$\frac{\partial E}{\partial z_0}=0$、$\frac{\partial E}{\partial k}=0$， then：

$$\begin{cases} \frac{\partial E}{\partial x_0}=2\sum_{i=1}^{n}(x_i^2-2x_ix_0+z_i^2-2z_iz_0+k)(-2x_i)=0 \\ \frac{\partial E}{\partial z_0}=2\sum_{i=1}^{n}(x_i^2-2x_ix_0+z_i^2-2z_iz_0+k)(-2z_i)=0 \\ \frac{\partial E}{\partial k}=2\sum_{i=1}^{n}(x_i^2-2x_ix_0+z_i^2-2z_iz_0+k)=0 \end{cases} \qquad (9)$$

where $n$ is the number of contour points. By solving the equations, the diameter value of each circular projection is obtained with the pixel unit. Here we define the fitted diameter value as $W_i^{'}$, so we get the corresponding voxel size value:

$$V_i=U\frac{W_i^{'}}{W} \qquad (10)$$

where $W$ is the actual diameter of the spherical phantom which can be accurately measured by mechanical or photoelectrical means such as laser or coordinate measuring machine (CMM). Inputting the discrete values of $V$ and $D$ to equation (6), an equation set is got:

$$\begin{bmatrix} D_1 & 1 \\ D_2 & 1 \\ ..... & 1 \\ D_i & 1 \\ ..... & 1 \\ D_n & 1 \end{bmatrix}\begin{bmatrix} k \\ b \end{bmatrix}=\begin{bmatrix} V_1 \\ V_2 \\ ... \\ V_i \\ ... \\ V_n \end{bmatrix} \qquad (11)$$

After solving the equations (11), the least-square solutions are $b$ and $k$. Thereby the expression of the equation (6) is obtained. In line with the equation, as long as importing the movement distance of the rotary table, the corresponding voxel size is exported, thus to realize automatic calibration of the voxel size of the 3D-CT scanning system.

# 3. Results

In order to verify the effectiveness of the present calibration method, we used a steel ball with diameter of 19mm and fixed it on the support frame installed in the rotary table of the 3D-CT scanning system. The main configuration parameters of the 3D-CT scanning system are:

Size of X-ray focus: 5 $\mu$m

Size of imaging area: 200 *mm*×200 *mm*

Pixel size: 200 $\mu$m

Source to Detector Distance (SDD): 695 *mm*

In the experiment, we first moved the rotary table to the nominal zero position $P_0$ along Y-direction and captured a projection of the steel ball. 64 frames were averaged to one frame so as to increase the SNR(Signal to Noise Ratio) of the projection image. Next the rotary table was translated to its maximum distance position $P_{\max}$ along Y-direction and another projection of the steel ball was captured. Then the rotary table was moved to different positions between $P_0$ and $P_{\max}$ and several DR projections of the steel ball were acquired. After this, the image segmentation, edge detection, image thinning and moment match tracing are applied to each circular projection to obtain one closed circle contour. Fig. 4 shows one DR image of the steel ball and its edge contour after performing image processing when $D_i$=330mm. According to coordinates of the contours, the non-linear least square fitting method was employed to get the diameter of each circular projection in the pixel unit. The calculation results are shown in Tab.1. Meanwhile, we plotted $V_i - D_i$ graph as shown in Fig.5, from which we can find that a satisfying linear relationship between moving distance of the rotary table and the voxel size exits. Finally, the discrete couple values $[D_1, D_2, \ldots D_i \ldots D_n]$ and $[V_1, V_2, \ldots V_i \ldots V_n]$ were input into equation (11), its least square solutions were resulted in: $V = 0.00028698D\text{-}0.0346$, which was loaded into the calibration module of the 3D-CT scanning system to realize automatic calibration of the voxel size.

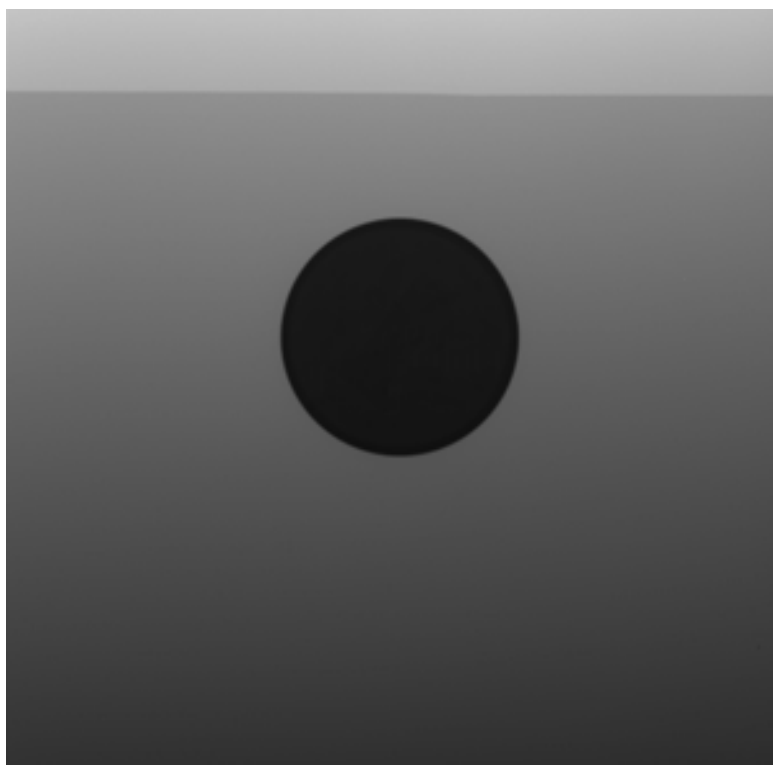
(a)

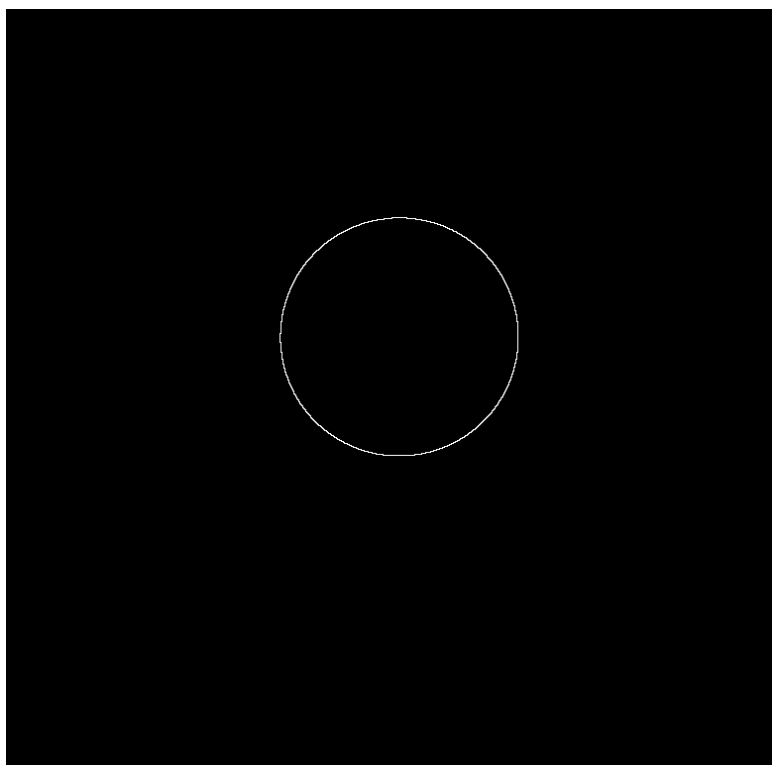
(b)

**Fig.4 DR projection and contour of a steel ball**

**Tab.1 The measurement results of $D_i$-$V_i$**

| $D_i$ | **Diameter** $W_i^{'}$ **(pixel)** | **GMR** | **Voxel size** $V_i$ **(×10$^{-2}$mm)** |
|---|---|---|---|
| 220 | 668.65 | 7.038 | 2.8412 |
| 250 | 512.25 | 5.392 | 3.7091 |
| 270 | 443.52 | 4.668 | 4.2841 |
| 300 | 368.75 | 3.882 | 5.1525 |
| 330 | 316.16 | 3.328 | 6.0098 |
| 350 | 284.28 | 2.993 | 6.6831 |
| 380 | 255.32 | 2.687 | 7.4422 |
| 400 | 237.10 | 2.496 | 8.0135 |
| 430 | 214.23 | 2.255 | 8.8681 |
| 450 | 201.28 | 2.119 | 9.4386 |

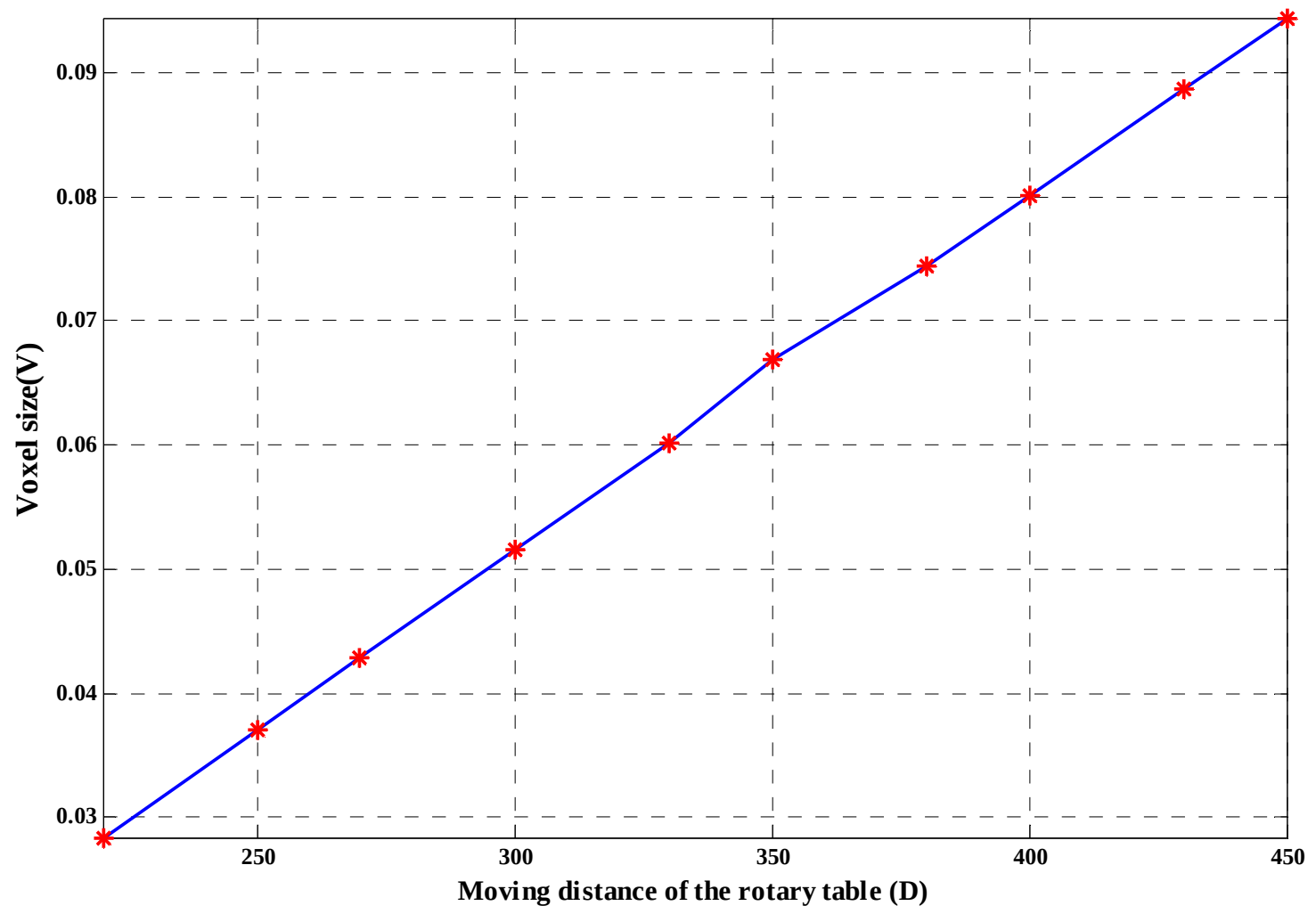

**Fig.5 Relationship diagram of** $D_i - V_i$

In order to verify the accuracy of the proposed method, we scanned a standard phantom in the same 3D-CT scanning system , then loaded all the cross-sectional images and the voxel size which was calculated by the proposed method into the 3D visualization and analysis software Mimics10 (Materialise，Belgium) as shown in Fig.6. On the base of the virtual 3D model ,we selected several typical dimension parameters indicated in Fig.6 and measured their value by the software. Tab.2 is the measurement results. We also measured the typical dimension parameters by CMM with the size measurement precision of 1~3μm. Comparing the two kinds of measurement results, we can find that the proposed method here provides a voxel size with high precision, which ensures the image analysis and feature measurement more accurate and credible.

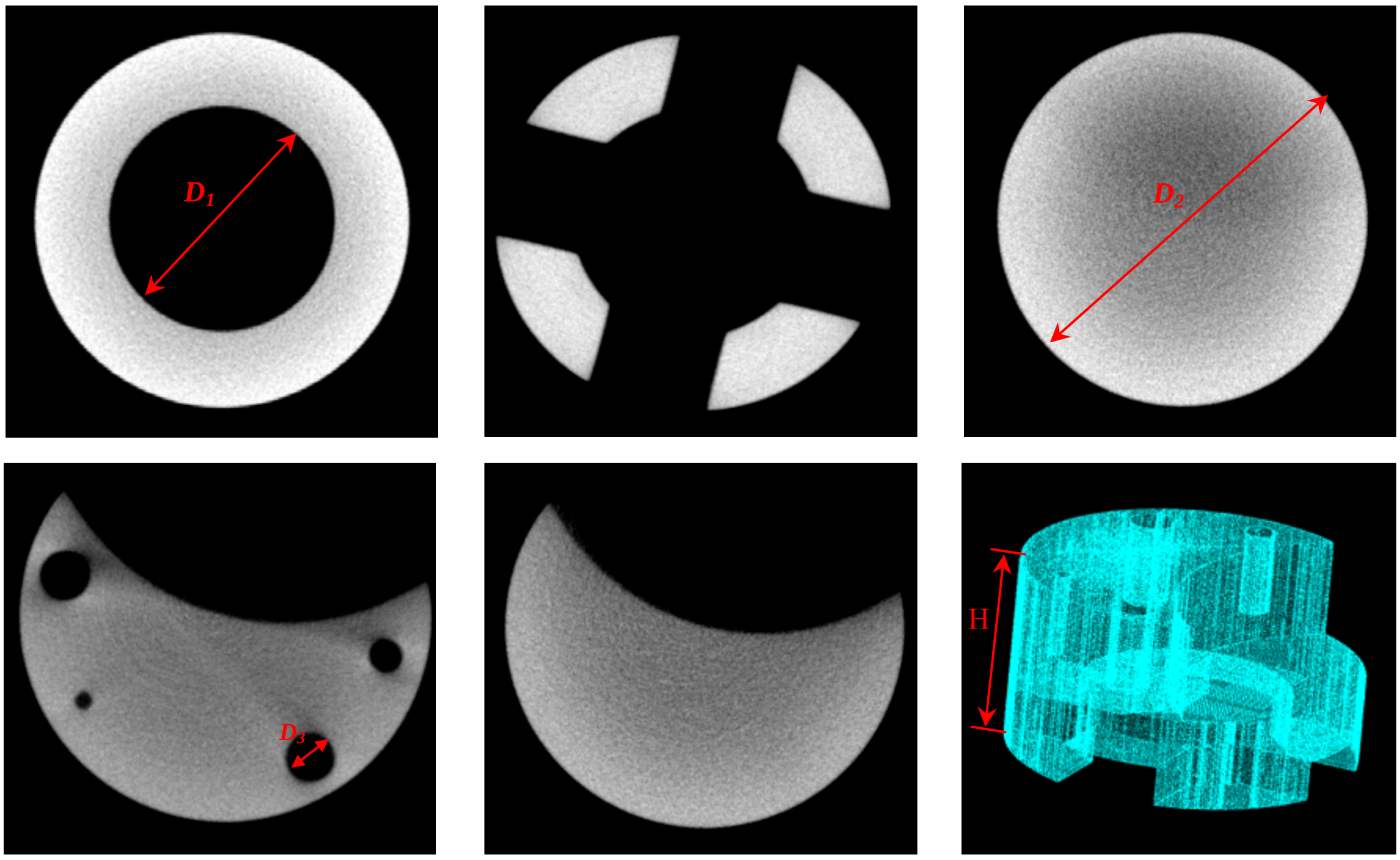

**Fig.6 CT slices and 3D virtual model of a standard phantom**

**Tab.2 Size measurement results of a standard phantom**

| Measurement | CMM (mm) | Mimics(mm) | Absolute error（mm） | Relative error (%) |
|---|---|---|---|---|
| $D_1$ | 44.9922 | 45.0861 | 0.0939 | 0.21 |
| $D_2$ | 74.9781 | 75.1512 | 0.0731 | 0.23 |
| $D_3$ | 9.0556 | 9.0934 | 0.0378 | 0.42 |
| $H$ | 49.9725 | 50.0403 | 0.0678 | 0.14 |

# 4. Conclusion

Voxel size is an important indicator to describe the resolution of a cone-beam 3D-CT scanning system and its accuracy also guarantees the precision of data analysis and feature measurement based on 3D-CT images. In this paper, we proposed an automatic calibration method to get the voxel size with high precision. In line with this method, several circular projections of a spherical phantom are captured at different imaging positions along Y-direction, where the corresponding moving distances are recorded by motor controller at the same time. Then image segmentation, edge detection, image thinning and moment match tracing are applied to each circular projection to obtain one closed circle contour. Based on the coordinates of the contour points, least square fitting is used to calculate the diameter of the circular projection. Through these interpolation values, a linear equation is obtained, which reflects the relationship between the rotary table displacement distance from its nominal zero position and the voxel size. This linear equation is finally imported into calibrating module of the 3D-CT scanning system, and when the rotary table is moving along Y-direction, accurate value of the voxel size is dynamically exported. The experiments results prove that this method meets the requirements of the actual 3D-CT scanning system, and has virtues of easy implementation and high accuracy.

**Acknowledgements : This work was supported in part by the National Natural Science Foundation of China (NSFC) under Grants 11275019，21106158 and 61077011, in part by the National State Key Laboratory of Multiphase Complex Systems under Grant MPCS-2011-D-03, in part by the National Key Technology R&D Program of China under Grant 2011BAI02B02, in part by the Beijing Municipal Commission of Education (BMCE) under the joint-building project.**

# 锥束 3D-CT 扫描系统重建体素尺寸自动标定方法

**杨民[1],王晓龙[1],刘义鹏[2],孟凡勇[3],李兴东[4],刘文丽[4],魏东波[1]**

1、 北京航空航天大学机械工程及自动化学院，北京 100191，中国
2、 哈尔滨工业大学控制科学与工程系，哈尔滨 150001, 中国
3、 中国科学院过程工程研究所 多相复杂系统国家重点实验室，北京 100190
4、 中国计量科学研究院，北京 100013, 中国

**摘要**：对于锥束 3D-CT（Cone-beam Three Dimensional Computed Tomography）扫描系统，重建体素尺寸是反映系统重建精度的一个重要参数，其精度决定着基于 CT 图像的特征分析与尺寸测量的精度，并且在成像过程中重建体素尺寸随着成像位置改变而动态变化。为了实现对重建体素尺寸的动态标定，利用球形目标体在不同几何成像放大比下的 DR（Digital Radiography）图像，配合图像、图形处理方法及最小二乘拟合技术求取不同成像位置下的目标体投影直径及对应位置的重建体素尺寸，然后回归出扫描转台的行程距离变量与体素尺寸变量之间的关系方程，从而实现锥束 3D-CT 扫描系统体素尺寸的动态自动标定。实验结果表明，该方法的测量精度达到了实际扫描系统的重建精度要求，并且实现简单。